# Mechanically Controlled Quantum Interference in Graphene Break Junctions

*Sabina Caneva[1], Pascal Gehring[1], Víctor M. García-Suárez[2,3], Amador García-Fuente[2], Davide Stefani[1] Ignacio J. Olavarria-Contreras[1], Jaime Ferrer[2,3,*], Cees Dekker[1] and Herre S. J. van der Zant[1,*]*

[1]Kavli Institute of Nanoscience, Delft University of Technology, Lorentzweg 1, 2628 CJ Delft, The Netherlands.
[2]Departamento de Física, Universidad de Oviedo, 33007 Oviedo, Spain.
[3]Nanomaterials and Nanotechnology Research Center, CSIC – Universidad de Oviedo, 33007 Oviedo, Spain

[*]Corresponding authors: H.S.J.vanderzant@tudelft.nl
ferrer@uniovi.es

**The ability to detect and distinguish quantum interference signatures is important for both fundamental research and for the realization of devices including electron resonators[1], interferometers[2] and interference-based spin filters[3]. Consistent with the principles of subwavelength optics, the wave nature of electrons can give rise to various types of interference effects[4], such as Fabry-Pérot resonances[5], Fano resonances[6] and the Aharonov-Bohm effect[7]. Quantum-interference conductance oscillations[8] have indeed been predicted for multiwall carbon nanotube shuttles and telescopes, and arise from atomic-scale displacements between the inner and outer tubes[9,10]. Previous theoretical work on graphene bilayers indicates that these systems may display similar interference features as a function of the relative position of the two sheets[11,12]. Experimental verification is, however, still lacking. Graphene nanoconstrictions represent an ideal model system to study quantum transport phenomena[13–15] due to the electronic coherence[16] and the transverse confinement of the carriers[17]. Here, we demonstrate the fabrication of bowtie-shaped nanoconstrictions with mechanically controlled break junctions (MCBJs) made from a single layer of graphene. We find that their electrical conductance displays pronounced oscillations at room temperature, with amplitudes that modulate over an order of magnitude as a function of sub-nanometer displacements. Surprisingly, the oscillations exhibit a period larger than the graphene lattice constant. Charge-**

**transport calculations show that the periodicity originates from a combination of quantum-interference and lattice-commensuration effects of two graphene layers that slide across each other. Our results provide direct experimental observation of Fabry-Pérot-like interference of electron waves that are partially reflected/transmitted at the edges of the graphene bilayer overlap region.**

We use mechanically controlled break junctions (MCBJs)[18] made from single-layer graphene to measure the electrical conductance of the junction during uniaxial deformation. The MCBJ device consists of a lithographically-defined graphene bridge supported on a flexible metal substrate. The substrate is bent, inducing stretching of the graphene bridge until it ruptures. The bending direction can be reversed, causing the graphene edges to approach each other such that electrical contact can be re-established and thereby allowing multiple ($>10^3$) measurements to be performed on a single junction.

Figures 1a-d show the nanofabrication process for the devices: (1a) Graphene transfer onto a polymer-coated metal substrate; (1b) Patterning and etching of the graphene bowties (constriction width of 400 nm and length of 750 nm); (1c) Evaporation of Ti/Au leads (1d); Removal of the polymer regions surrounding the bowtie. The sample is mounted in a 3-point bending configuration (1e) with a central pushing rod beneath the sample and two counter supports on top of the substrate at both ends. Fig. 1f shows a scanning electron microscope image of a device before performing the measurements, with a magnified view of the junction area in which the graphene bowtie is supported by the polymer structure beneath it. Without the support structure, the edges of the broken junction cannot reform a contact upon unbending (see Fig. S1).

A typical experiment consists of breaking the graphene bridge at room temperature in air by gradually bending the substrate until a sharp decrease in conductance is observed, dropping from an initial conductance of ~1 $G_0$, down to the noise level of the setup (~1×10$^{-6}$ $G_0$). Here, $G_0$ is the quantum of conductance defined as $2e^2/h$ = 77 µS. For a detailed discussion of the conductance calibration see Fig. S2. We fabricated twelve junctions that displayed a conductance >1 $G_0$ before bending: these were used for the MCBJ experiments and four of which exhibited clear conductance oscillations. Examples of first breaking traces for different samples (C, D, E) are shown in Fig. 2a. Our results are in line with conductance values reported in the literature for monolayer graphene flakes exfoliated in situ with a conductive tip, which are close to 1 $G_0$ immediately prior to rupture[19].

After the graphene bridge is broken, the electrode gap is closed by unbending the substrate (i.e. electrical contact restored) and can be re-opened by bending the substrate; consecutive breaking-making traces are thus recorded as a function of time and displacement of the electrodes. A striking feature of these traces is the consistent conductance recovery from ~$1\times10^{-6}$ $G_0$ to ~1 $G_0$ in each cycle, indicating excellent mechanical stability and reproducibility of the measurement (Fig. 2b,c). Remarkably, large periodic oscillations appear in the high-conductance range, with drops of almost an order of magnitude between maxima. These features are similar to the current modulations observed in suspended graphene bilayers during indentation with an AFM tip, albeit with much larger amplitudes[20]. These oscillations are not observed for the first breaking trace and only appear during all following cyclings of the junction. In order to convert the time axis units into horizontal displacement, we record 20 *I-V*s for fixed displacements in the conductance regime from the noise level up to ~$1\times10^{-4}$ $G_0$ for this sample. Given that the *I-V*s in this range display tunnelling characteristics, the curves are fitted with a Simmons model[21] from which the gap size can be estimated (see Fig. S3 and Fig. S4). Figure 2d shows the alignment of 1,000 consecutive breaking measurements as a function of displacement for Sample A. The close similarity of all the traces is evident from the vertical lines along the displacement axis, which shows that these oscillations are not merely random noise but exhibit a defined periodicity. A two-dimensional (2D) histogram is constructed from the aligned breaking traces (Fig. 2e), clearly showing the same modulations present in the individual curves. The corresponding alignment and 2D histogram for the making traces are shown in Fig. S5.

Figure 3a shows representative breaking traces for two different samples that display oscillations. For each sample, 1,000 traces are summed and the derivative with respect to the distance is calculated to determine the periodicity of the signal. Figure 3b shows the linear fit for the peak number as a function of peak position (peaks are identified by the maxima and the minima in the derivative), from which we estimate a typical oscillation periodicity of 0.68 nm for Sample A (blue curve) and 0.89 nm for Sample B (magenta curve).

Figure 3c displays the average of 10 breaking traces at room temperature in air and vacuum ($10^{-6}$ mbar) for Sample F. We observe that the period is increased from 0.60 nm to 0.68 nm going from air to vacuum (Fig.

3d). To relate this change to a shift in the Fermi wave vector, $k_F$, and therefore doping level, we fabricated similar graphene bowtie structures on SiO$_2$(285 nm)/Si and measured the source-drain current as a function of gate voltage for analogous conditions. For the sample in the inset of Fig. 3d, when exposed initially to air, the Dirac point ($V_{Dirac}$) is at ~70 V. In vacuum $V_{Dirac}$ shifts to ~50 V. We ascribe the higher $V_{Dirac}$ in air to the well-known *p*-type doping due to the adsorption of oxygen and water molecules during ambient exposure[6]. From the Fabry-Pérot interference condition $dk_F = \pi$, we can compare the experimentally derived periodicity ratio $d_{vac}/d_{air}$=1.33 with the ratio $\sqrt{V_{Dirac,air}/V_{Dirac,vac}}$ = 1.84 ± 0.61 as determined from measuring seven gated samples (see Table S1). Hence, the periodicity falls within this range. The $\sqrt{V_{Dirac,air}/V_{Dirac,vac}}$ values are extracted given $k_F = \sqrt{n\pi}$, and $n = C_{ox}(V_{gate}-V_{Dirac})/e$, where $n$ is the carrier concentration, $C_{ox}$ is the back gate capacitance, and setting $V_{gate}$ to zero (no gate voltage is applied in the MCBJ experiments). Our findings can be understood from the fact that the less doped graphene in vacuum has a wave-vector which is smaller than the wave-vector for the air-exposed sample. Using the Fabry-Pérot interference condition, this would result in larger oscillation periodicities for the vacuum-exposed sample, which is in line with our experiments. We can thus exclude the lattice periodicity effect as the only origin of the conductance oscillations. The inset in Fig. 3d shows a schematic of the low energy dispersion relation for *p*-type doped graphene and the relative change in the wave-vector with doping level.

To unveil the physical origin of the observed conductance oscillations, we perform density functional theory simulations of two graphene sheets sliding rigidly against each other and calculate the conductance for each relative displacement. Figure 4a shows simulated breaking/making cycles where two sheets are moved relative to each other in the zigzag direction. Given that the experiments are performed in ambient conditions, the sheets are terminated at the contact region by armchair edges that are either passivated by hydrogen atoms (blue curve) or carboxyl groups (magenta curve). The shape of these conductance traces agrees remarkably well with the experiments; the traces start at ~1×10$^{-6}$ G$_0$, saturate at around 0.5 G$_0$ and show conductance oscillations of very similar shape and period. These quantum interference oscillations should appear if the Fabry-Pérot interference condition $k_F d_{QI} = \pi$ is fulfilled[20], where $k_F = 2\pi/3c$ is the Fermi wave vector along the zigzag direction, $d_{QI}$ is the Fabry-Pérot period and $c$ = 0.246 nm is the graphene lattice constant. The oscillations' period in the zigzag direction should therefore be $d_{QI} = 3/2\ c = 0.37$ nm. These oscillations cannot be seen directly in Fig. 4a because they overlap with those originating from the lattice periodicity effect, which have a period equal to $c$ in the zigzag direction (see Figure S11 and Table S2).

Figure 4b allows the two types of oscillations to be disentangled by plotting the magnitude of the oscillations' amplitude, measured as the ratio between the maximum and minimum value of the conductance of each oscillation ($G_{max}/G_{min}$) in Fig. 4a, as a function of the distance between two corresponding minima. We find that our data points cluster into three groups. The first group has a conductance ratio of ~1-2, and an oscillation period equal to $c$, which is much smaller than the periods measured in our experiments. These oscillations originate from the lattice periodicity effect. The amplitudes of the second group of oscillations are larger, with a conductance ratio of ~10. This group clusters around $d_{QI} = 0.37$ nm and corresponds to Fabry-Pérot interferences. The third group clusters at a period of 0.74 nm = $3c \approx 2\ d_{QI}$. Here, the two shorter oscillations combine, which results in a beating pattern with larger oscillation period and very big amplitudes with conductance ratios of ~100. Conductance traces for bilayers sliding along the armchair and chiral (4,1) directions are discussed in the Supplementary Information (see Figs. S6-S10). The overall picture that emerges is coherent: the length scale of the oscillations due to the lattice periodicity effect is of the order of $c$ or a fraction of it, e.g.: much smaller than the period measured in our experiments. In contrast, Fabry-Pérot oscillations, as well as the combined effect of the Fabry-Pérot and potential oscillations, have longer periods of at least 0.4 nm and even reaching 1 nm, which are consistent with our experimental findings.

Figure 4c shows a contour plot of the electron transmission through the junction $T$, as a function of the energy of the impinging electrons $E$ and the displacement $d$. The plot shows both oscillations and interference patterns. A first set of oscillations becomes periodic and independent of energy for sufficiently negative $d$ (i.e., for increased bilayer overlap region), with a period equal to the lattice spacing in the zigzag direction. These oscillations originate from the periodic nature of the hopping integrals that bridge the two sheets, which modify the sheet band structure. We find a second set of oscillation lines that runs diagonally and bends towards the bottom left in the plot (i.e., large bilayer overlap and negative energy). These are caused by the destructive interference of waves partially confined in the bilayer region.

Figure 4d shows the calculated transmission contour plot as a function of $E$ and $d$ based on the one-dimensional tight binding model in Fig. 4e, where both vertical and curved oscillations are clearly discernible, resulting from changes in the lattice commensuration and anti-resonances respectively. We model the bilayer system as two oppositely oriented semi-infinite chains that slide on top of each other. An algebraic analysis of the model shows that electron waves are partly reflected/transmitted at the edges of the chain overlap region of length $D = (N+1)c \approx d$, where $N$ and $c$ are the number of atoms in this overlap region and the lattice constant, respectively. This results in destructive interference whenever $k_F (N+1)c = n\pi$, where $k_F$ is the Fermi wave vector and $n$ is a positive integer. The periodic vertical oscillations in the contour vanish when we set the inter-chain hopping integrals $t_{ij}$ to be constant (i.e. removing the shifts in the lattice commensuration), however, the curved interference pattern remains. Additional one-dimensional models and further algebraic details that support this analysis are described in the Supplementary Information (see Fig. S14 and Fig. S13). These results are in good agreement with previous theoretical studies on the conductance variation in bilayer graphene as a function of overlap length[22].

In conclusion, we presented graphene mechanically controlled break junctions, and demonstrated that the junction conductance can be reversibly switched by almost six orders of magnitude during 1,000 opening-closing cycles, attesting to the high mechanical stability of the MCBJ device. While electronic switching has been demonstrated for electroburned graphene nanoribbons, the switching behaviour was attributed to the formation/rupture of carbon chains[23,24]. We additionally observed large conductance oscillations as a function of nanometer displacement between two overlapping graphene sheets during cycling. This is in contrast to the featureless conductance traces in twisted graphene tunnelling junctions reported in the literature[25]. Our findings, supported by density functional theory and tight binding calculations, are a direct experimental observation of quantum interference of standing waves in sliding bilayer graphene at room temperature.

Our MCBJ platform allows the intriguing possibility of tuning the junction conductance over a large range with a purely mechanical tuning knob, thus finding potential use as a nanoelectromechanical system[20,26]. Furthermore, given that the gap size can be adjusted with exceptionally high precision and stability, we envisage that graphene MCBJs can be used as a powerful tool for quantum tunnelling-based sensing of (bio)molecules[27,28].

**Acknowledgements**

S.C. acknowledges a Marie Skłodowska-Curie Individual Fellowship under grant BioGraphING (ID: 798851) and P.G. acknowledges a Marie Skłodowska-Curie Individual Fellowship under grant TherSpinMol (ID: 748642) from the European Union's Horizon 2020 research and innovation programme. This work was supported by the Graphene Flagship (a European Union's Horizon 2020 research and innovation programme under grant agreement No. 649953), the Marie Curie ITN MOLESCO and an ERC advanced grant (Mols@Mols No. 240299). The research by V.M.G.S., A.G. and J.F. was funded by the project FIS2015-63918-R from the Spanish government.

**Author Contributions**

S.C., H.S.J.Z. and C.D. conceived the idea and designed the experiments. S.C. developed the nanofabrication protocol. S.C., I.J.O.C., D.S. and P.G. performed the break junction experiments. P.G. and S.C. performed the graphene gating measurements. P.G. designed and implemented the cross correlation method and performed the *I-V* data analysis. J.F. supervised the theoretical research work. V.M.G.S. and J.F. conceived the simulations. A.G.F. and V.M.G.S. carried out the DFT calculations. V.M.G.S. and J.F. carried out the tight binding calculations. J.F. developed the algebraic analysis of the charge transport model and of the interference conditions. All authors participated in discussions and co-wrote the paper.

**Competing interests**

The authors declare no competing financial interests.

**Additional information**

Supplementary information is available in the online version of the paper. Reprints and permissions information is available online at www.nature.com/reprints. Correspondence and requests for materials should be addressed to H.S.J.Z. and correspondence regarding the theoretical calculations should be addressed to J.F.

**Figure Captions**

**Fig. 1. Nanofabrication of graphene mechanically controlled break junctions. a)** Transfer of monolayer graphene on top of the phosphor bronze substrate coated with polyimide (PI) and polymethylglutarimide (PMGI) insulating layers. **b)** Reactive ion etching of the graphene into a bowtie shape by $O_2$ plasma (400 nm constriction width and 750 nm Au electrode separation). **c)** Evaporation of leads [Ti(5 nm)]/Au(70 nm)]. **d)** Development of the sample in ethyl lactate to produce a supported graphene bridge. **e)** Schematic representation of the MCBJ measurement setup. **f)** False-colour scanning electron microscope image of a typical device before performing the measurements and magnified view of the junction.

**Fig. 2. Electromechanical measurements reveal conductance oscillations. a)** Conductance, plotted on a logarithmic scale, as a function of time for the first breaking trace in Samples C, D and E measured at a bias voltage of 0.1 V. **b)** Example of typical breaking and **c)** making trace as a function of displacement and time during cycling of Sample A, showing periodic oscillations in the high-conductance range ($10^0$-$10^{-4}$ $G_0$, where $G_0$ is the quantum unit of conductance). The arrows indicate the direction of the measurements. We define the zero displacement ($d = 0$) as the position in which the last C atom of the left graphene edge is located directly beneath the last C atom of the right graphene edge. From density functional theory calculations such a configuration yields a conductance between $10^{-2}$-$10^{-4}$ $G_0$. **d)** Alignment of 1,000 traces using a cross correlation method in which each trace is shifted such that the overlap with all other traces is maximised. **e)** Two-dimensional histogram of conductance as a function of displacement constructed from 1,000 consecutive traces from Sample A measured at room temperature in air. No trace selection was performed. For all data presented in the main text, the contact resistance has been subtracted from the measured conductances (see Fig. S2).

**Fig. 3. Analysis of the oscillation periodicities. a)** Typical breaking trace for Samples A and B displaying oscillations. The vertical dotted line indicates the displacement, *d,* at which the graphene contact is broken, defined as $d = 0$. **b)** Peak position plotted as a function of peak number for the first 6 and 7 peaks in Sample A and B respectively. The experimental data points are shown as blue and magenta circles. The line of best

fit provides an estimate of the oscillation periodicity: 0.68 nm for Sample A and 0.89 nm for Sample B. **c)** Comparison of the oscillations in the $10^{-1}$ $G_0$ to $10^{-3}$ $G_0$ conductance regime for Sample F exposed to air and vacuum ($10^{-6}$ mbar) at room temperature. We recorded 100 breaking traces for each experiment. Given the measured conductance range, we cannot define the same $d = 0$ reference as in panels **a)** and **b)**. We have thus arbitrarily set the zero at the start of the traces. Inset: source-drain current as a function of gate voltage showing Dirac curves for a graphene bowtie on $SiO_2$/Si exposed to air and vacuum ($10^{-5}$ mbar vacuum with a mild annealing step, see Methods). **d)** Peak position as a function of peak number (labelled 1-5 in panel (c)), showing a larger period for the vacuum-exposed sample. Inset: schematic of the energy-momentum relation, *E-k*, depicting the change in the Fermi wave vector, $k_F$, with doping level.

**Fig. 4. Density functional theory and tight-binding calculations of sliding graphene bilayers. a)** Examples of two calculated breaking traces for graphene sheets sliding along the zigzag direction and terminated with armchair edges**.** The conductance is plotted on a logarithmic scale as a function of the displacement, *d*. The edges have been passivated by hydrogen (blue) and carboxyl groups (magenta) and the intersheet distance is ~0.45 nm. **b)** Plot of the ratio of the maximum and minimum value of the conductance of the oscillations, $G_{max}/G_{min}$, as a function of the distance between adjacent minima normalized by the lattice constant *c* for each of the oscillations in Fig. 4a. The labels $d_P$, $d_{QI}$ and $d_{QI+P}$ denote the periodicity due to the lattice periodicity effect, to quantum interference effects and to the combination of the two effects respectively. The oscillation with the largest amplitude has a period $d_{QI+P} \sim 0.74$ nm. **c)** Contour plots of the decimal logarithm of the transmission function $\log_{10}(T(E,d))$ as a function of the displacement *d* and of the energy *E* referenced to the Fermi energy $E_F$ for the hydrogen-terminated bilayer in Fig 4a. **d)** Contour plots of $\log_{10}(T(E,d))$ of the model in c), as a function of *d* and *E* where parameters have been adjusted to fit the experimental distances, graphene hopping integrals and tunneling decay lengths. The plot displays the different types of conductance oscillations: a vertical periodic pattern due to shifts in lattice commensuration (white areas) and destructive quantum interferences (dark blue features). These are also apparent in c). **e)** Schematic of the tight-binding model showing the relevant inter-chain ($r_0$) and inter-atomic ($r_{ij}$) distances, the chain ($\varepsilon_0$) and edge on-site ($\varepsilon_1$) energies, and intra-chain ($t_0$) and inter-chain ($t_{ij}$) hopping parameters. $D = (N+1)c \approx d$ is the overlap length, and $N$ and $c$ are the number of atoms in this overlap region and the lattice constant, respectively.

**Methods**

**Nanofabrication of graphene MCBJs**

The MCBJ devices were prepared using the following protocol, based on the work of Rickhaus *et al.*[29]. Flexible phosphor bronze (PB) chips (1×2 cm$^2$) were cleaned by sonicating 5 min in acetone and subsequently IPA. An adhesion promoter (VM651) was spin coated on the PB at 2500 rpm and baked 1 min at 110 °C. An insulating layer of polyimide (PI) was spin coated on top at 800 rpm and baked in a vacuum oven at 300 °C for 30 min. A layer of lift-off resist of (SF7) was spin coated as the topmost layer at 4500 rpm and baked 5 min at 180 °C. A PMMA resist layer (A6 495k) was spin coated at 4500 rpm and baked at 180 °C for 2 min. Markers were patterned on the chip by e-beam lithography (Raith EBPG 5000+). The sample was developed in xylene at room temperature for 30 s and then blow dried in N$_2$. A 60 nm thick W layer was sputtered and lift-off was performed in xylene at 80 °C for 10 min, followed by blow drying in N$_2$. PMMA-covered monolayer CVD graphene (purchased from Graphenea on Cu foil) was transferred using the wet etching process in ammonium persulfate. The PMMA was removed by placing the sample in xylene at 80 °C for 10 min. A new layer of PMMA (A6 495k) was spin coated at 4500 rpm and baked at 180 °C for 2 min. The graphene was patterned into bowtie shapes by exposing the region around it (electron dose of 850 μC/cm$^2$). The sample was developed in xylene at room temperature for 30 s. The exposed graphene regions were etched by reactive ion etching in an O$_2$ plasma (5 mbar, 20 W, 20 sccm, 30 s). The PMMA was removed by placing the sample in xylene at 80 °C for 10 min. A double layer resist (PMMA A6 495k and A3 950k) was spin coated at 4500 rpm and baked at 180 °C for 2 min after each layer. Leads and pads were patterned (electron dose of 850 μC/cm$^2$) and the sample was developed in xylene at room temperature for 30 s. E-beam evaporation was used to deposit Ti(5 nm)/Au(70 nm). Lift-off was performed in xylene at 80 °C for 30 min. Lastly, the sample was dipped in ethyl lactate for 30 s to produce an undercut and then rinsed twice in DI water.

**MCBJ experiments**

Large Au pads connecting the graphene bowties were covered in silver conducting paste and the chip was mounted in a custom-made break junction set-up. The conductance was recorded as a function of time at a bias voltage of 0.1 V while a brushless servo motor bent the sample. Once the conductance dropped to the noise level, the motor was stopped. A voltage was then applied on the piezo element to verify whether the junction could be cycled (i.e., recover a high-$G$ state by decreasing the piezo voltage and breaking once more by applying a voltage). Once cycling was confirmed, a conductance histogram measurement comprising 1,000 consecutive traces was set up. All experimental plots display the conductance of the graphene junction only (i.e., the contact resistance is subtracted by measuring in a 4-point probe scheme). Measurements were performed at room temperature in air and in vacuum ($10^{-6}$ mbar).

**Measurements with back-gate**

For the determination of the effect of doping between air and vacuum measurements, graphene bowtie samples on silicon backgate substrates ($SiO_2$(285 nm)/Si) were prepared based on a previously published protocol[6]. The samples were measured in air and in vacuum ($10^{-5}$ mbar) using home-built low-noise DC electronics). The gate voltage was swept from -20 V to 90 V, while a bias of 10 mV was applied between source and drain electrodes. To account for the very high current per unit width on the order of 0.5-10 A/mm (using $G = G_0$ and a contact width of about 1-20 nm) that can arise in very narrow constrictions formed in the graphene break junction experiments, a current of 100 µA (which corresponds to a current per unit width of about 0.5 A/mm in the constriction area) was passed through the junction for several minutes in vacuum.

**Density Functional Theory simulations**

Ab initio simulations of graphene MCBJs have been carried out with the DFT program SIESTA[30]. Here, two sheets oriented along zigzag, armchair or chiral directions are terminated with armchair, zigzag or complementary chiral edges. The edges are passivated by hydrogen or carboxyl groups. Closing MCBJ cycles are simulated where the two sheets approach each other in steps of 0.05 Ångström, starting from a distance where the tunnelling rate is zero. The conductance and electron transmission calculations have been performed with the quantum transport code GOLLUM[31]. Further details as well as sketches of the MCBJ geometries of the simulations can be found in the Supplementary Information (see Figs. S6-S13).

The DFT code SIESTA is publicly available at https://launchpad.net/siesta

**Tight Binding modelling**

The electrical transport features of the several one-dimensional models have been analysed algebraically and/or numerically as discussed in the main text and in the Supplementary Information (see Fig. S12 and Fig. S13).

GOLLUM is a quantum transport package that is freely available upon request at http://www.physics.lancs.ac.uk/gollum/

**Data Availability**

The data that support the plots within this paper and other findings of this study are available from the corresponding author upon reasonable request.

**Figure 1**

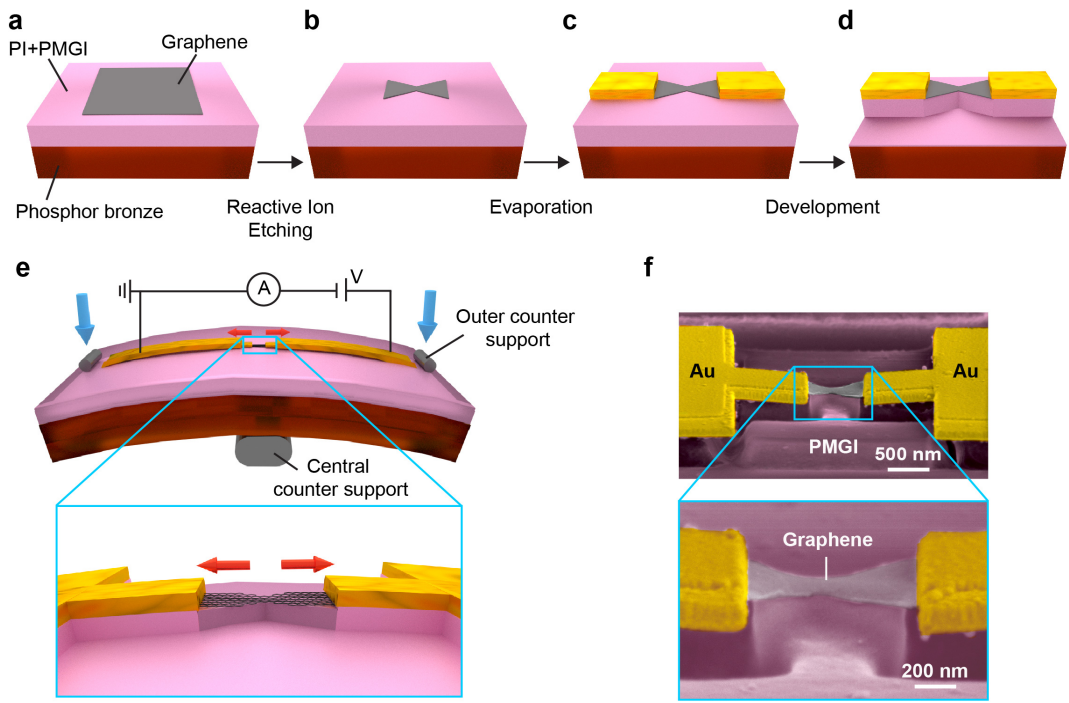

**Figure 2**

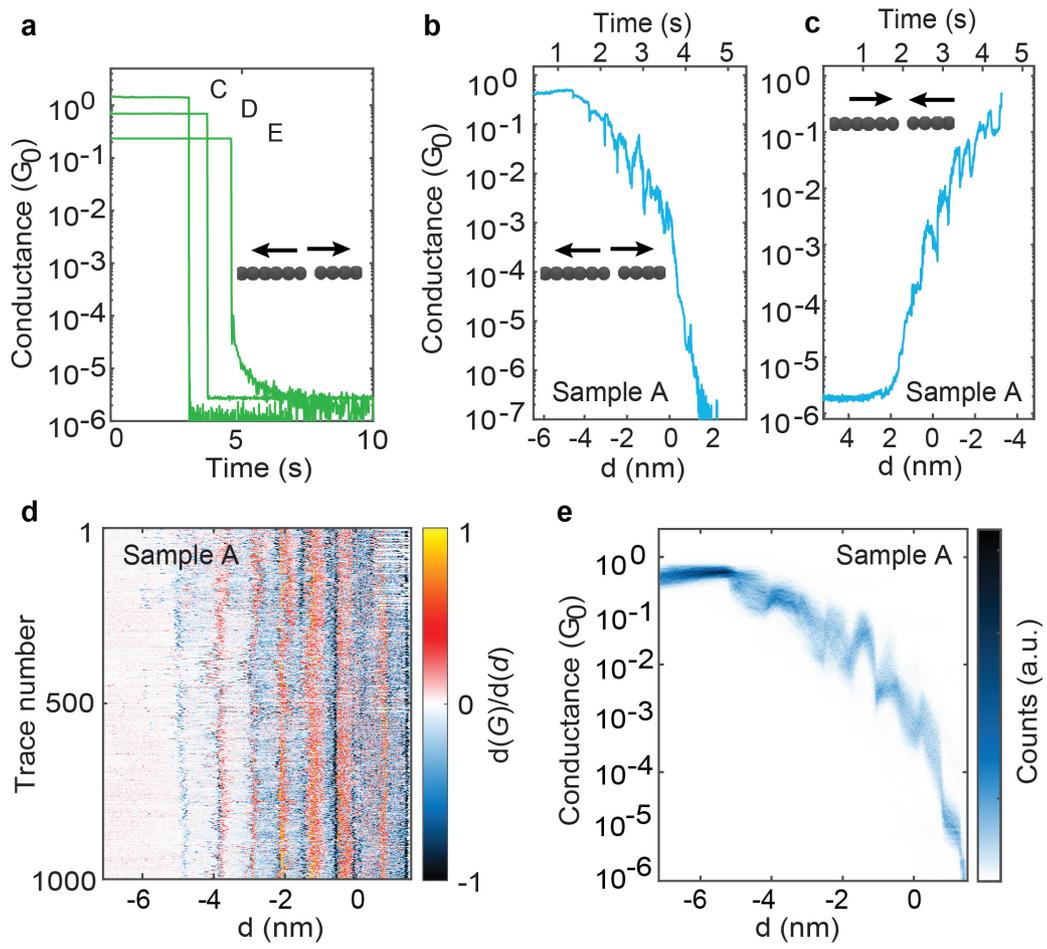

**Figure 3**

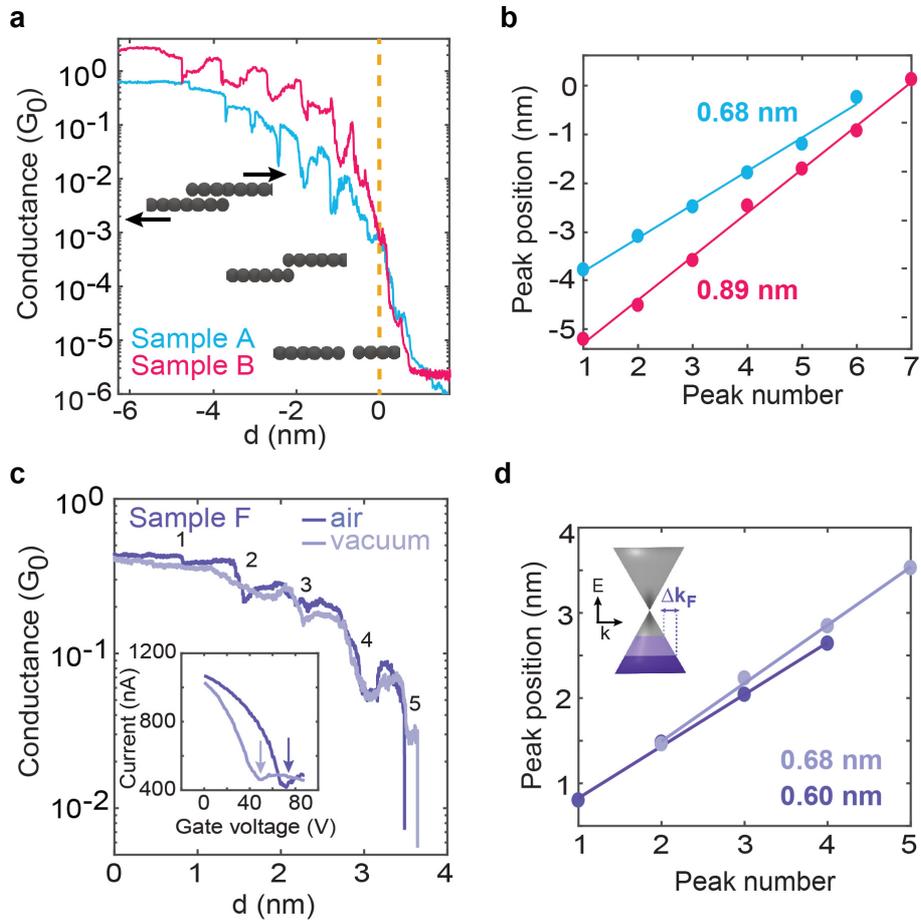

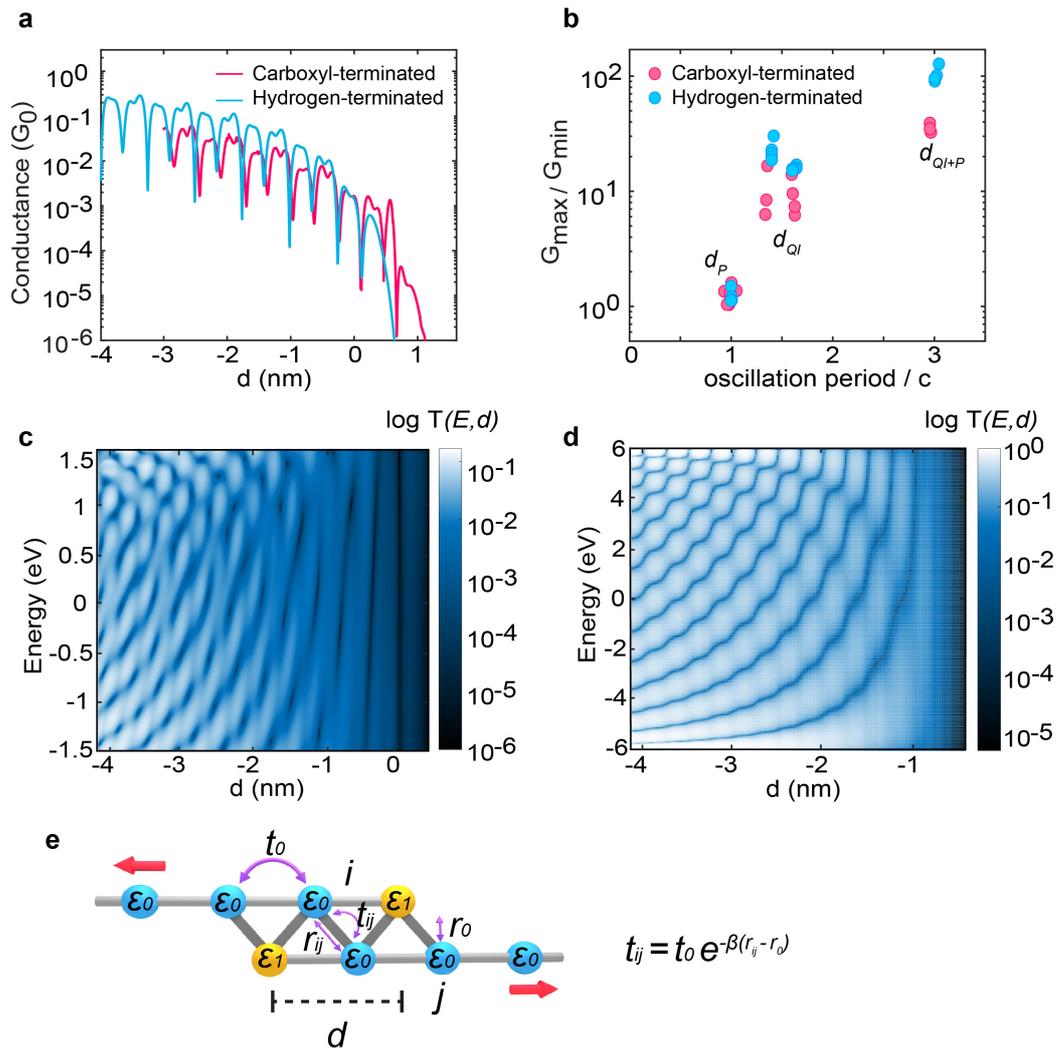